\documentclass{emulateapj}

\def\ms{m~s$^{-1}$}
\def\mjup{M$_{\rm Jup}$}

\def\msun{M$_{\odot}$}
\def\msini{$M_P\sin i~$}
\def\chisq{$\sqrt{\chi^2_\nu}$}
\def\chis{$\chi^2_\nu$}

\def\feh{[Fe/H]}

\def\mkvk{$\{V-K$, $M_{K}\}$}

\def\starA{Gl\,649}
\def\spt{M1.5}
\def\dA{10.34}
\def\udA{0.15}
\def\mpA{0.328}
\def\umpA{0.032}
\def\feA{$+0.08$}
\def\ufeA{0.06}
\def\pA{598.3}
\def\upA{4.2}
\def\pyearsA{1.638}
\def\upyearsA{0.011}
\def\eA{0.30}
\def\ueA{0.08}
\def\kA{12.4}
\def\ukA{1.1}
\def\tpA{12876}
\def\utpA{22}
\def\omA{352}
\def\uomA{15}
\def\rmsA{4.2}
\def\chiA{1.17}
\def\mstarA{0.54}
\def\umstarA{0.05}
\def\arelA{1.135}
\def\uarelA{0.035}
\def\nobs{44}

\begin{document}
\title{The California Planet Survey II. A Saturn-Mass Planet Orbiting the M Dwarf Gl\,649$^1$} 

\author{John Asher Johnson\altaffilmark{2},
Andrew W. Howard\altaffilmark{3,4},
Geoffrey W. Marcy\altaffilmark{3},
Brendan P. Bowler\altaffilmark{5},
Gregory W. Henry\altaffilmark{6},
Debra A. Fischer\altaffilmark{7},
Kevin Apps\altaffilmark{8},
Howard Isaacson\altaffilmark{3},
Jason T. Wright\altaffilmark{9}
}

\email{johnjohn@astro.berkeley.edu}

\altaffiltext{1}{ Based on observations obtained at the
W.M. Keck Observatory, which is operated jointly by the
University of California and the California Institute of
Technology. Keck time has been granted by both NASA and
the University of California.}
\altaffiltext{2}{Department of Astrophysics,
  California Institute of Technology, MC 249-17, Pasadena, CA 91125}
\altaffiltext{3}{Department of Astronomy, University of California,
Mail Code 3411, Berkeley, CA 94720}
\altaffiltext{4}{Townes Fellow, Space Sciences Laboratory, University of
  California, Berkeley, CA 94720-7450, USA} 
\altaffiltext{5}{Institute for Astronomy, University of Hawai`i, 2680
  Woodlawn Drive, Honolulu, HI 96822} 
\altaffiltext{6}{Center of Excellence in Information Systems, Tennessee
  State University, 3500 John A. Merritt Blvd., Box 9501, Nashville, TN 37209}
\altaffiltext{7}{Yale Astronomy Department, Box 208101, New Haven, CT 06520-8101, USA}
\altaffiltext{8}{Cheyne Walk Observatory, Horley, Surrey, RH6 7LR, United Kingdom} 
\altaffiltext{9}{The Pennsylvania State University, University Park,
  PA 16802}

\begin{abstract}
We report precise Doppler measurements of the nearby ($d = \dA$~pc) M
dwarf \starA\ that reveal 
the presence of a planet with a minimum mass \msini~$ =
$~\mpA~\mjup\ in an eccentric ($e = \eA$), \pA~day orbit. Our
photometric monitoring reveals \starA\ to be a new variable star with
brightness changes on   
both rotational and decadal timescales.  However, neither of these 
timescales are consistent with the 600-day Doppler signal
and so provide strong support for planetary reflex motion as the best
interpretation of the observed radial velocity variations. \starA\,b is
only the seventh Doppler-detected giant planet around an M dwarf.
The properties of the planet and host-star therefore contribute 
significant information to our knowledge of planet formation around
low-mass stars. We revise and refine the occurrence rate of giant
planets around M dwarfs based on the California Planet Survey sample
of low-mass stars ($M_\star < 0.6$~\msun). We find that $f =
3.4^{+2.2}_{-0.9}$\% of stars with 
$M_\star < 0.6$~\msun\ harbor planets with \msini~$>
0.3$~\mjup\ and $a < 2.5$~AU. When we restrict our analysis to
metal-rich stars with [Fe/H]~$ > +0.2$ we find the occurrence
rate is $10.7^{+5.9}_{-4.2}$\%. 
\end{abstract}


\section{Introduction}

Compared to the knowledge gleaned from the large sample
of giant planets around Sun-like stars, little is known about the
characteristics of Jovian planets around M dwarfs. This is due 
primarily to the empirical finding that the occurrence rate of
detectable 
planets scales with stellar mass \citep{johnson07b}; low-mass stars
($M_\star < 0.6$~\msun) 
simply do not harbor giant planets very frequently
\citep{endl03,butler06b}. The frequency of giant
planets with \msini~$ > 0.3$~\mjup\ around Sun-like stars is 
8\% within 2.5 AU \citep{cumming08}, and the occurrence of giant planets
around M dwarfs is roughly a factor of 4 lower. 

\begin{figure}[t!]
\epsscale{1.2}
\plotone{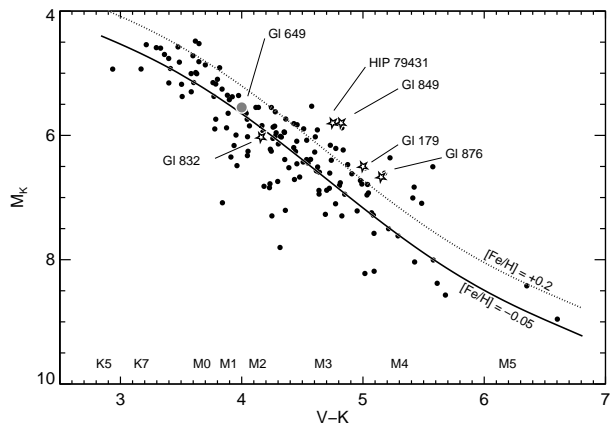}
\caption{Our Keck sample of low-mass stars plotted in the
  \mkvk\ plane. The solid line is a fifth-order polynomial fit to the
  mean main sequence for stars within 10~pc, which Johnson \& Apps
  (2009) identify as an isometallicity contoure with [Fe/H] equal to
  the mean value of the Solar neighborhood. The dashed line
  corresponds to \feh~$ = +0.2$ based on the calibration of Johnson \& 
  Apps. The five-point stars show the positions of all of the M
  dwarfs known to harbor at least one giant planet. The solid circle
  denotes the position of \starA.
 \label{fig:hr}} 
\end{figure}

While the lower masses of M dwarfs decreases the likelihood of
giant planet occurrence, a handful of Jovian planets have been
discovered around low-mass stars. The sample of M dwarfs known
to harbor at least one Doppler-detected giant planet (\msini~$ >
0.2$~\mjup) is listed in Table~\ref{tab:mplanets} and shown in the
H--R diagram in Figure~\ref{fig:hr}. Also given in that table
are the stellar and planetary masses from the literature, and stellar
metallicities from the broad-band photometric calibration of 
\citet{johnson09b}.    

\begin{figure*}[ht!]
\epsscale{1.0}
\plotone{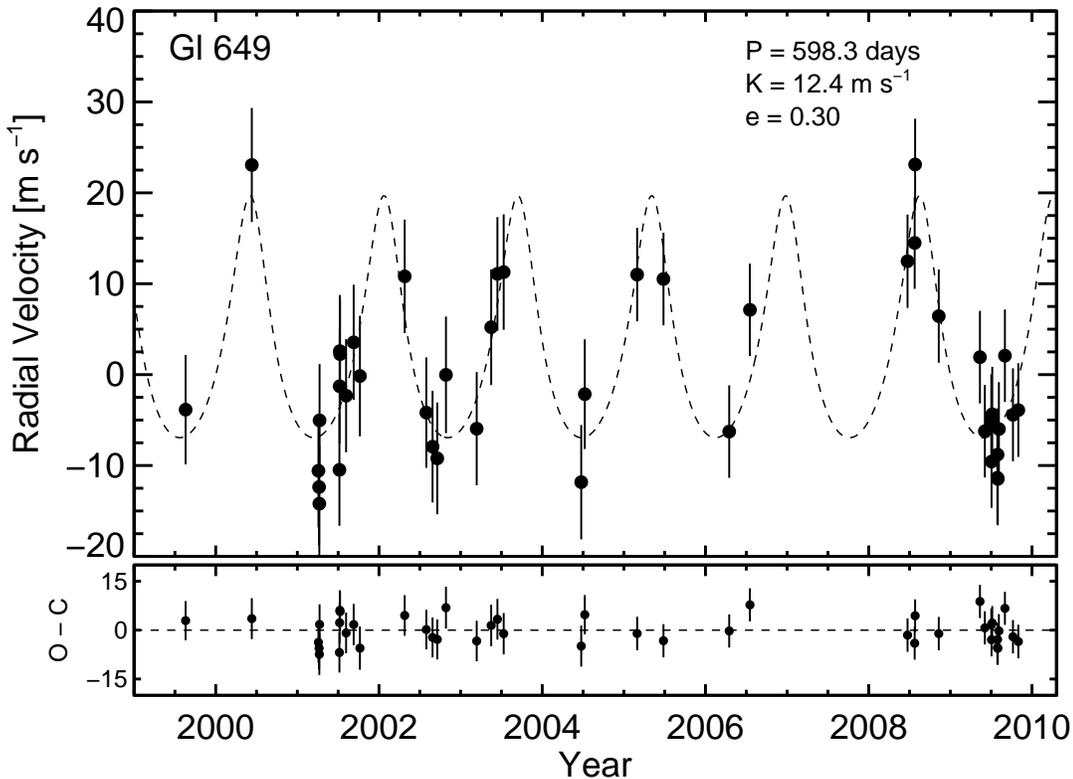}
\caption{
Our Keck/HIRES radial velocity time series for \starA. The dashed line
shows the  best--fitting Keplerian model. The rms scatter of the
residuals (bottom panel) is \rmsA~\ms, and \chisq~$ = \chiA$. 
 \label{fig:rv}} 
\end{figure*}

These planets and their host stars demonstrate that stellar mass is
not the only characteristic that correlates with the probability of a
star harboring a planet. Stellar metallicity has been shown to be a strong
predictor of planet occurrence around Sun-like stars \citep{fv05}, and the
correlation between planet frequency and stellar metal content appears
to hold for the M dwarfs, as well. \citet{johnson09b} found that M
dwarfs with Jovian planets tend to be significantly metal-rich
compared to a 10 pc, volume-limited sample of stars on the lower main
sequence. For example, Gl\,849 harbors a Jovian planet in a
long-period orbit and is among the most metal-rich stars in
the Solar neighborhood with [Fe/H]~$ > +0.45$. 

If this preliminary
trend proves to be real then it will provide valuable constraints for
theoretical models of planet formation around a broad range of stellar
characteristics. The effect of metallicity on planet occurrence will
also inform the target selection of future Doppler and transit surveys
targeting low-mass stars \citep[e.g.][]{irwin08}, as well as
the interpretation of results from direct-imaging, astrometric and
microlensing surveys \citep[e.g.][]{nielsen09, pravdo09, dong09}. 

As the time baselines, sample sizes and Doppler precision increase for
the various 
Doppler surveys of low-mass stars, the relationships between the
physical characteristics of stars and the properties of their planets
will come into sharper focus. We are 
monitoring a sample of 147 late K and early M stars as part of the
California Planet Survey at Keck Observatory 
with a current temporal baseline of $\approx12$~years and Doppler
precision of 2--3~\ms\ (Johnson et al. 2007;
Howard et al. 2009b). In this contribution we announce the detection
of a new Saturn-mass planet orbiting a nearby M dwarf. \starA\ is only
the eleventh M-type star 
known to harbor at least one Doppler-detected
planet\footnote{Several additional low--mass host stars have been 
discovered by gravitational microlensing surveys \citep[e.g.]{bond04,
  gould06,  beaulieu06, dong09}.}, and it is only
the seventh low-mass star with a Doppler-detected giant planet
\citep[see also][for examples of low-mass planet
detections]{bonfils05b,maness07,forveille09,mayor09b}. In the following 
section we describe the stellar properties of \starA, and our
spectroscopic observations and Doppler-shift measurements.
In \S~\ref{sec:null} we test the validity of our interpretation of the
observed radial velocity variations by measuring the false-alarm
probability and by examining our photometric measurements. We conclude in 
\S~\ref{sec:discussion} with a summary and discussion of \starA\,b,
and we place this latest exoplanet in context with 
other giant planets discovered around M dwarfs.

\section{Observations and Analysis}

\subsection{Stellar Properties}

\starA\ ($=$~HIP\,83043) is an \spt\ dwarf with a {\it Hipparcos}
parallax-based distance of $10.34 \pm 0.15$ parsecs 
 \citep{hipp2}, apparent magnitude $V= 9.7$, and absolute magnitude
 $M_V = 9.627 \pm 
 0.053$ \citep{hipp}. We use the broadband metallicity calibration of 
\citet{johnson09b} to estimate [Fe/H]~$ = \feA \pm 0.06$, and we adopt the
stellar mass estimate provided by the \citet{delfosse00} K$_s$-band
mass-luminosity 
relationship, which gives $M_\star = \mstarA \pm 0.05$~\msun. Using the
infrared flux method, \citet{alanso96} give an effective temperature
$T_{\rm eff} = 3700 \pm 60$~K. \citet{wright04b} measured the emission
in the CaII\,H 
emission line relative to the stellar photosphere on the Mt. Wilson
scale \citep{duncan91} and give a median ``grand $S$'' value of 
$1.55$. This $S$ value places the chromospheric activity of
\starA\  among the top 20\% of nearby early M-type 
stars, as shown by \citet{rauscher06} and by \citet{gizis02}.
Our spectra show H-alpha to be in absorption, as was found for all
Balmer lines observed in the spectrum of \starA\ \citep{gizis02}.  The 
stellar properties of \starA\ are summarized in Table~\ref{tab:starorbit}. 

\subsection{Radial Velocities and Keplerian Fit}

We began monitoring \starA\ at Keck Observatory in 1999 October using
the High-Resolution Echelle  
spectrometer \citep{vogt94} in our standard Iodine cell setup
with the B5 decker, giving a reciprocal resolution
$\lambda/\Delta\lambda = 55,000$ per $\sim 4$-pixel resolution element 
\citep{ howard09}. We measured the Doppler shifts of the star 
from each star-times-iodine observation using the standard analysis
procedure presented by \citet{butler96}, with subsequent improvements
over the years. For HIRES observations made prior to
the 2004 CCD upgrade the measurement uncertainties range from
3.3--4.4~\ms, and improve to 0.9--1.5~\ms\ thereafter.

\begin{figure}[t!]
\epsscale{1.2}
\plotone{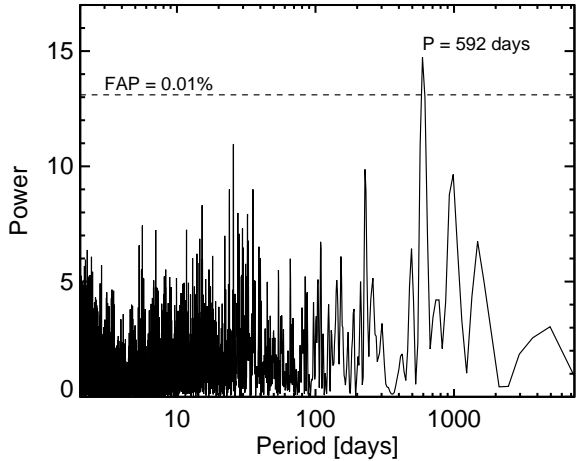}
\caption{Periodogram analysis of our RV time series, which reveals a
  strong peak at $P = 592$~days. The dashed lines show the analytic
  false-alarm probability of a peak arrising from noise sampled at our
  times of observation.
 \label{fig:pergram}} 
\end{figure}

Our \nobs\ radial velocities are presented in Table~\ref{tab:rv}
(without jitter) and the time series is shown in Figure~\ref{fig:rv}
(with jitter). The scatter in the  
measurements is larger 
than expected from the measurement errors, and a periodogram analysis
of the data reveals strong power at periods near 592 days, with a
corresponding analytic false-alarm probability $< 0.0001$
(Figure~\ref{fig:pergram}).  

We used the partially-linearized Keplerian fitting code
\texttt{RVLIN}\footnote{http://exoplanets.org/code/} 
described by \citet{wrighthoward} to search for a best-fitting orbital
solution to the data. To ensure proper weighting of our measurements in the
fitting procedure, we inflated the error bars to account for RV noise
from astrophysical sources. This stellar ``jitter'' term is calculated
based on the star's chromospheric activity, $B-V$ color and absolute
V-band magnitude using the formula of \citet{wright05}. We adopt a 
jitter estimate of 3~\ms\ for \starA, which we add in quadrature to the
measurement errors. 

We find that a single-planet Keplerian model with a period $P =
\pA \pm \upA$~days, eccentricity $e = \eA \pm \ueA$ and velocity semiamplitude
$K = \kA \pm \ukA$~\ms\ results in a root-mean-squared (rms) scatter of
\rmsA~\ms\ in 
the residuals and \chisq~$ = \chiA$, indicating an
acceptable fit\footnote{We use \chisq\ to indicate the factor by which
the observed scatter about the best-fitting model differs from our
expectation based on the measurement errors. Thus, the 
scatter about our model is a factor of \chiA\ larger than our
average error bar.}. The resulting minimum 
planet mass is \msini~$=\mpA$~\mjup, and the semimajor axis is $a =
\arelA$~AU. The best-fitting solution is shown in Figure~\ref{fig:rv},
with the residuals to the fit shown in the lower panel. The orbital
parameters are listed in Table~\ref{tab:starorbit}. 

The parameter uncertainties given above were estimated using a Markov
Chain Monte Carlo (MCMC) algorithm with $10^7$ links, in which a
single randomly-chosen 
parameter was perturbed at each link, with a perturbation size tuned
such that 20--40\% of the jumps were executed \citep[see e.g.][ and
  references therein]{ford05, winn08b}. The resulting ``chains'' of
parameters form the posterior probability distribution, 
from which we select the 15.9 and 84.2
percentile levels in the cumulative distributions (CDF) as the
``one-sigma'' confidence limits. In most cases the posterior
probability distributions were approximately Gaussian.

\section{Testing the Null--Hypothesis}
\label{sec:null}

\subsection{False-Alarm Probability}

The Doppler semi-amplitude of our
best-fitting model, $K=$~\kA~\ms, is 
comparable to the measurement uncertainties and stellar
jitter, which prompted us to test the null--hypothesis that the
apparent 
periodicity arose by chance from larger-than-expected radial velocity
fluctuations and sparse sampling. We tested this possibility
calculating the false-alarm probability (FAP) based on the goodness of
fit statistic $\Delta\chi^2_\nu$ 
\citep{howard09,marcy05a,cumming04}, which is the difference between
two values of $\chi^2_\nu$: one from the single-planet Keplerian fit and
one from the fit of a linear trend to the data. Each trial is
constructed by keeping the times of observation fixed and scrambling
the measurements, with replacement.  We record the
$\Delta\chi^2_\nu$  value after each trial and repeat this process for
10,000 trial data sets. For the ensemble set we compare the resulting
distribution of $\Delta\chi^2_\nu$ to the value from the fit to the
original data. 

We found that none of the $10^4$ trials resulted in a
higher value of $\Delta\chi^2_\nu$, which we interpret as a  $ <
0.0001$ probability that the 600-day periodicity is a spurious signal
due to random fluctuations.

\subsection{Photometric Variability}
\label{sec:phot}

\begin{figure}[t!]
\epsscale{1.2}
\plotone{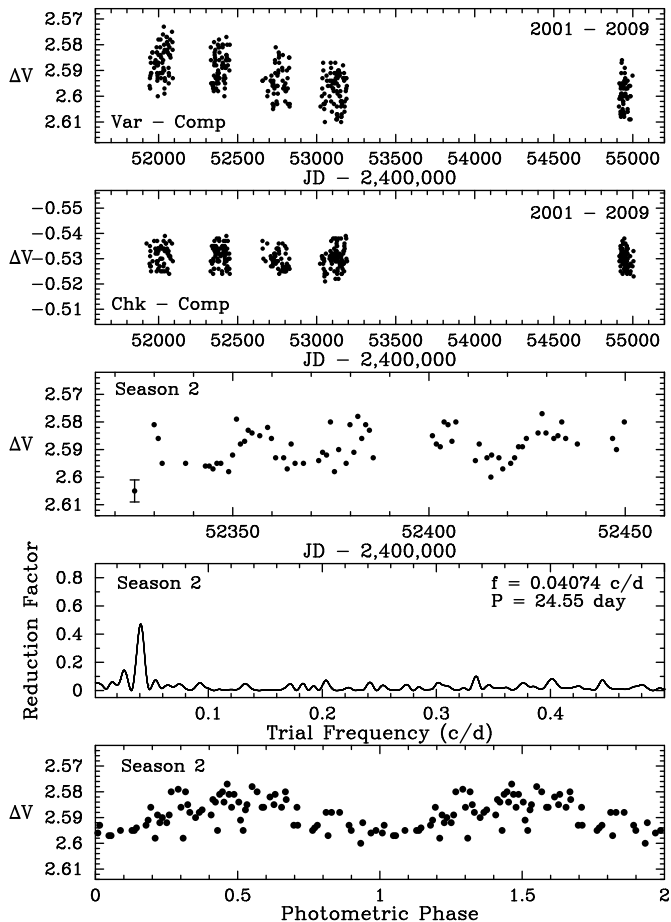}
\caption{$Top~Panel$:  Our 337 $V-C$ photometric observations of Gl\,649 
in the Johnson $V$ band acquired with the T3 0.4m APT at Fairborn 
Observatory.  $Second~Panel$:  The $K-C$ observations plotted with an
identical scale as the top panel.  Comparison of these two data sets 
shows that Gl 649 varies in brightness on night-to-night and year-to-year
timescales.  $Third~Panel$:  Observations from the second observing season
plotted on an expanded x axis clearly show low-amplitude brightness
variability in Gl 649.  $Fourth~Panel$:  Frequency spectrum of the
observations from season two gives a best period of 24.55 days.  
$Bottom~Panel$:  Plot of the data from season two phased with the 
24.55-day period reveals coherent variability with a peak-to-peak
brightness amplitude 0.012 mag. \label{fig:phot}} 
\end{figure}

We note that our FAP value only addresses the existence of a
periodicity in the radial velocities, but does not test its
cause. As an additional test of the null hypothesis we acquired
brightness measurements of \starA\ in the Johnson $V$ 
passband with the T3 0.4~m automatic photometric telescope (APT) at 
Fairborn Observatory.  The APT observations cover five observing
seasons between 2001 February and 2009 June and reveal photometric
variability in Gl 649 on both rotational and decadal timescales.  Details 
on APT operations, data acquisition and reduction procedures, and
precision of the observations can be found in 
\citet{henry95,henry95b,fekel05,eaton03}.  

Our 337 \starA--minus--comparison ($V-C$) differential magnitudes are 
plotted against heliocentric Julian Date in the top panel of
Figure~\ref{fig:phot}.   
The comparison star is HD\,152342 ($V=7.10$, $B-V=0.35$, F2V). Most obvious 
in this plot are the year-to-year changes in the mean magnitude of the 
observations.  The mean magnitudes have a range of 0.0126 mag and suggest 
the possible existence of a spot (magnetic) cycle in Gl 649 with a length 
of at least several years \citep[see, e.g.,][]{henry99,hall09}.  The 
top panel also shows that the range in the $V-C$ observations is 
$\sim$0.02 mag within all five observing seasons.  The standard deviations 
of the individual five seasons are all between 0.0057~mag and 0.0063~mag.

The check minus comparison ($K-C$) differential magnitudes are plotted in 
the second panel of Figure~\ref{fig:phot} at the same magnitude scale
as the top panel.   
The check star is HD\,153897 ($V=6.57$, $B-V=0.43$, F4V).  The yearly means 
of the $K-C$ measurements have a range of only 0.0016 mag; their standard 
deviation is only 0.0007 mag.  The standard deviations of the individual
five seasons are all between 0.0034 mag and 0.0042 mag.  Thus, the larger 
night-to-night scatter of the $V-C$ measurements and the observed 
year-to-year change in the $V-C$ mean magnitudes must both be intrinsic 
to \starA.  

The observations from season~2 are replotted in the third panel of 
Figure~\ref{fig:phot}, again on the same magnitude scale as in panels 1 and 2.  
Low-amplitude variability is clearly seen with a period of 20--30 days. 
Panel~4 shows our computed frequency spectrum for season~2, where the 
y axis plots the reduction factor in the variance of the observations for 
each trial frequency \citep{vanicek71}.  We find a clear period of 
24.55 days; the observations are plotted phased with this period in the
bottom panel of Figure~\ref{fig:phot}.  A least-squares sine fit to the
phase curve 
gives a peak-to-peak amplitude of $0.0117\pm0.0015$ mag. Similar analyses 
yield periods of 23.72, 27.92, 25.80, and 21.86 days for seasons one, 
three, four, and five, respectively.  The mean of these five periods is 
$24.8\pm1.0$ days, which we take to be the rotation period of Gl 649
revealed by rotational modulation in the visibility of cool starspots
on the photosphere of Gl 649.  This photometric variability is consistent 
with the level of chromospheric activity in the star, as mentioned
above. 

\citet{henry95} show many examples of active stars with low-amplitude 
starspot variability. \citet{queloz01b} and \citet{paulson04} show several 
examples of stellar spots masquerading as planets.  In the case of Gl\,649
described here, the photometric observations reveal variability timescales
that are inconsistent with the 600-day radial velocity
variations. Sinilarly, we examined the time-variability chromospheric
emission from each of our spectroscopic observations. While the
$S$-value, as measured from the Ca\,II~H\&K emission, has a variance
of 0.14 dex, we observed no peridicities near the putative orbital
period. The lack of photometric 
and chromospheric variability provide additional strong support for the
interpretation of planetary reflex 
motion as the cause of the observed radial velocity variability in
Gl\,649.  

\section{Discussion}
\label{sec:discussion}
We have presented the discovery of a Saturn-mass planet (\msini~$ =
\mpA$~\mjup) orbiting the nearby, low-mass star \starA\ ($d = $~\dA~pc,
\mstarA~\msun). \starA\,b resides in an eccentric ($e = \eA$) orbit
with a period of \pA~days, corresponding to a  semimajor axis $a =
\arelA$~AU. 

\starA\ is only the seventh M dwarf known with a Doppler-detected
giant planet, and the fifth detection from among the 147 low-mass stars
we've monitored over the past decade at Keck Observatory
\citep[e.g.][]{johnson07b}. The low Doppler amplitude of the planet ($K =
\kA$~\ms) highlights our need to attain high measurement
precision to find low-mass planets, and to maintain that precision
over long time baselines to detect planets at larger
semimajor axes.  

\citet{johnson07b} recently analyzed the detection rate among our
sample of low-mass stars and reported a 1.8\% occurrence rate of
planets with $a < 2.5$~AU. The increased time baseline and new
detections of our sample suggest that a reanalysis of the frequency of
planets around M dwarfs is warranted. Following Johnson et al. we
first note that the $\approx10$~year time baseline of our survey,
together with our radial velocity 
precision, provides us with sensitivity to planets with \msini~$
\gtrsim 0.3$~\mjup\ out to semimajor axes $a \approx 2.5$~AU, assuming
and average stellar mass 
$M_\star = 0.5$~\msun. Note that in the analysis that follows, we exclude
the recently detected planets Gl\,832\,b, which was discovered by the
Anglo-Australian Observatory 
planet search \citep{bailey08}, and HIP\,79431\,b, which was only recently
added to the Keck survey as part of the metallicity-biased M-to-K
program (Apps et al. 2010, submitted). 

The probability density function (PDF) for the fraction of stars with 
planets given our number of detections $k =5$ and total sample
size $N = 147$ is given by the binomial distribution P$(f|k,N) \propto
f^5 (1-f)^{147-5}$. The overall occurrence 
rate from our sample is given by the maximum of the PDF, which we
measure to be 
$f = 3.4^{+2.2}_{-0.9}$\%, where the upper and lower limits represent the
68.3\% confidence interval measured from the cumulative distribution function. 

The corresponding giant planet fraction around
Sun-like stars was recently measured by \citet[][cf their Table
  1]{cumming08}, who report $f = 7.6 
\pm 1.3$\%. In a similar study Bowler et al. (2009, submitted)
measured the planet 
fraction around stars with $M_\star > 1.5$~\msun\ to be
$26^{+9}_{-8}$\%, albeit for minimum masses \msini~$ \gtrsim
1$~\mjup. Thus, the detection rate of giant planets around M dwarfs
consistently 
lags behind that of higher mass stars, despite the enhanced detectability
of planets around less massive stars since $K 
\propto M_\star^{-2/3}$ for a fixed planet mass and period. 
The contrast between the measured planet fractions between M dwarfs
and massive stars points to an even stronger correlation between
stellar mass and planet occurrence than measured by
\citet{johnson07b}. 

The correlation between stellar mass and planet formation is an
important piece of observational evidence in support of the core
accretion model of planet formation. In this model,
giant planets form in a bottom-up process, starting with the 
collisions of small dust grains and proceeding up through the
formation of large protoplanetary cores
\citep[see][for reviews]{ida04,alibert05}. Once these
cores attain a critical mass of $\sim10$~M$_\earth$, they can rapidly
accrete gas from the surrounding disk. Given the limited lifetime of
the gas disk, which dissipates on timescales less than 5~Myr
\citep{hernandez08,currie09b}, the formation of gas giant planets is a
race against time that is rarely won in the protoplanetary disks of
low-mass stars. The low density of raw materials, low orbital
frequencies ($\Omega \propto 1/P \propto M_\star^{1/2}$ at fixed $a$),
and unfavorable temperature 
profiles in the disks around M-type stars greatly inhibit the core
growth, which results in a lower occurrence of giant planets
\citep{laughlin04, ida05b, kennedy08, dr09}.  

Another important predictor of planet occurrence is stellar
metallicity. \citet{fischer05b} showed the fraction of Sun-like
stars with planets correlates strongly with [Fe/H], with an occurrence
rate of $\sim3$\% for \feh~$ < 0$ and a rise to $\approx25$\% for
\feh~$ > +0.3$. Until recently, it was difficult to properly account
for metallicity among the M dwarfs because the LTE spectral analysis
tools used for more massive stars are not amenable to the complex
spectra of low-mass stars \citep{maness07}. Because of the lack of knowledge
about the metallicity distribution of M dwarfs in general, and
low-mass stars with planets in particular, it was difficult to
determine whether stellar mass or metallicitiy lay at the root cause
of the puacity of planets around M dwarfs. 

The mass/metallicity issue was recently addressed by
\citet{johnson09b}, who derived a revised broad-band photometric
metallicity calibration for M dwarfs. They examined a sample of M
dwarfs with F, G and K wide binary companions. By anchoring the
metallicity of the M dwarf to its earlier-type companion, Johnson \&
Apps observed that metal-rich M stars reside ``above'' the mean
main sequence of the solar neighborhood when viewed in the
\mkvk\ plane. Further, they noticed the majority of the 7
planetary systems (containing planets of all masses) that were known
at the time contain metal-rich host stars.  

Figure~\ref{fig:hr} shows that with the addition of 3 new planet-host
stars since the study of Johnson \& Apps the planet-metallicity
correlation among M dwarfs appears to persist. We can quantify this
relationship by examining the fraction of stars in our Keck survey
with \feh~$ \geq 0$ that harbor giant planets. Including \starA, we
find that all 4 of the stars that harbor at least one giant planet\footnote{We
exclude Gl\,317 in this anlaysis because it lacks a reliable parallax
measurement and cannot be accurately placed in the \mkvk\ plane. The
existing parallaxes given in the literature suggest that Gl\,317
resides above the main sequence $\Delta M_K = $~0.08--0.37, indicating
either a Solar composition or that the star is metal-rich ($0.0 \lesssim
$~\feh~$\lesssim +0.15$).} 
fall within the subsamlple of 80 targets with \feh~$\geq 0$. Based on
this, we measure a planet fraction $f = 5.5^{+2.7}_{-2.1}$\% for
\feh~$ > 0$. If we restrict our analysis to \feh~$ > +0.2$ (dashed
line in Figure~\ref{fig:hr}), 3 of these 33 ``super-metal-rich'' stars
harbor planets, corresponding to $f = 10.7^{+5.9}_{-4.2}$\%. 

The uncertainties in our measured planet fractions are large due to
the small sample sizes involved. This underscores the need for
extending the time baseline of our current survey and expanding the
target list to include additional low-mass stars. Future surveys of
nearby, low-mass stars
such as the M2K planet search (Apps et al. 2010, submitted) and the
{\it MEarth} transit survey \citep{irwin09} will build upon the
current sample and provide a clearer picture of the planet-metallicity
relationship suggested from our analysis. A larger
sample of planets detected around M dwarfs will also provide crucial
leverage in understanding the relationship between stellar mass and
planet properties, especially when compared to the growing sample of
planets discovered around massive stars with $M_\star > 1.5$~\msun. 

\acknowledgments
We thank the many observers who contributed to the velocities reported here.  
We gratefully acknowledge the efforts and dedication of the Keck
Observatory staff, especially Grant Hill and Scott Dahm for support of
HIRES and Greg Wirth for support of remote observing. We are also
grateful to the time assignment committees of NASA, NOAO, and the
University of California for their generous allocations of observing
time. We acknowledge R.\ Paul Butler and S.\ S.\ Vogt for many years
of contributing to the data presented here.
A.\,W.\,H.\ gratefully acknowledges support from a Townes
Post-doctoral Fellowship at the U.\,C.\ Berkeley Space Sciences
Laboratory. J.\,A.\,J.\ thanks the NSF Astronomy and Astrophysics
Postdoctoral Fellowship program for support in the years leading to
the completion of this work, and acknowledges support form NSF grant
AST-0702821. G.\,W.\,M.\ acknowledges NASA grant NNX06AH52G.  
J.\,T.\,W.\ received support from NSF grant AST-0504874.
G.\,W.\,H acknowledges support from NASA, NSF,
Tennessee State University, and the State of Tennessee through its
Centers of Excellence program. 
Finally, the authors wish to extend special thanks to those of
Hawaiian ancestry  on whose sacred mountain of Mauna Kea we are
privileged to be guests.   
Without their generous hospitality, the Keck observations presented herein
would not have been possible.

\bibliography{}

\clearpage

\begin{deluxetable*}{cllllccl}
\tablecaption{Properties of M Dwarfs with Giant Planets \label{tab:mplanets}}
\tablewidth{0pt}
\tablehead{
  \colhead{Gliese} & 
  \colhead{{\it Hipparcos}} & 
  \colhead{Spectral} & 
  \colhead{Stellar}    &
  \colhead{[Fe/H]}   &
  \colhead{\msini~(\mjup)}    &
  \colhead{Semimajor} &
  \colhead{Reference}  \\
  \colhead{Number} & 
  \colhead{Number} & 
  \colhead{Type} & 
  \colhead{Mass (\msun)}    &
  \colhead{}   &
  \colhead{}    &
  \colhead{Axis (AU)} &
  \colhead{}
}
\startdata
876      & 
113020   & 
M4       & 
0.32     &
 $+0.37$ & 
0.6189, 1.9275\tablenotemark{a}  & 
0.207, 0.130   &
\citet{rivera05}       \\
849     & 
109388  &
M3      & 
0.45    & 
$+0.58$ & 
0.83\tablenotemark{b}    & 
2.35    &
\citet{butler06b}       \\
317     & 
...     & 
M4      & 
0.24    & 
...     & 
1.17    & 
0.95    &
\citet{johnson07b}       \\
832     & 
106440  &
M2      & 
0.45    &
$-0.12$ & 
0.64    & 
3.4     &
\citet{bailey09}       \\
179     & 
22627   &
M4      & 
0.36    & 
$+0.30$ & 
0.9     & 
2.42    &
Howard et al. (2009)  \\
...     & 
79431   & 
M3.5    & 
0.50    &
$+0.4$  & 
1.1     &
0.34    &
Apps et al. (2010, submitted)  \\
649     & 
83043   &
\spt    & 
\mstarA &
$+0.1$  & 
\mpA    &
\arelA  &
This work  \\
\enddata
\tablenotetext{a}{The Gl\,876 planetary system contains two resonant
  Jovian planets and an inner ``super-Earth'' with \msini~$=5.9$~$M_\earth$.}
\tablenotetext{b}{The orbit solution for Gl\,849 includes a linear
  velocity trend $dv/dt = -4.7$~m~s$^{-1}$~yr$^{-1}$, which may
  correspond to a second planet.}
\tablenotetext{c}{The orbit solution for Gl\,317 includes a linear
  velocity trend $dv/dt = 7.6$~m~s$^{-1}$~yr$^{-1}$, which may
  correspond to a second planet.}
\end{deluxetable*}

\begin{deluxetable*}{lc}
\tablecaption{Stellar Properties and Orbital Solution for \starA
\label{tab:starorbit}}
\tablewidth{0pt}
\tablehead{
\colhead{Parameter} & \colhead{Value} 
}
\startdata
$V$   & $9.70 \pm 0.04$ \\
$K$   & $5.62 \pm 0.02$ \\
$V-K$ & $4.08 \pm 0.05$ \\
$B-V$ & $1.52 \pm 0.04$ \\
$M_V$ & $9.63 \pm 0.05$ \\
$M_K$ & $5.55 \pm 0.02$ \\
$d$ (pc) & \dA~$ \pm $~\udA \\
$M_\star$ (\msun) & \mstarA~$ \pm $~\umstarA \\
T$_{\rm eff}$ (K) & $3700 \pm 60$ \\
\feh & \feA~$ \pm $~\ufeA \\
$P$ (days) &  \pA ~$ \pm $~\upA \\
$P$ (years) &  \pyearsA ~$ \pm $~\upyearsA \\
$K$ (m\,s$^{-1}$) & \kA~$ \pm $~\ukA \\
$e$ & \eA ~$ \pm $~\ueA\\
$T_P$ (Julian Date$ - 2400000$) & \tpA~$ \pm $~\utpA \\
$\omega$ (degrees) & \omA ~$ \pm $~\uomA \\
\msini\ (\mjup) & \mpA~$ \pm $~\umpA \\
$a$ (AU) & \arelA~$ \pm $~\uarelA \\ 
N$_{\rm obs}$ & \nobs \\
rms (m\,s$^{-1}$) & \rmsA \\
\chisq & \chiA \\
\chis & 1.37 \\
\enddata
\end{deluxetable*}

\begin{deluxetable}{lll}
\tablecaption{Radial Velocities for Gl\,649 \label{tab:rv}}
\tablewidth{0pt}
\tablehead{
\colhead{JD} &
\colhead{RV} &
\colhead{Uncertainty} \\
\colhead{-2440000} &
\colhead{(m~s$^{-1}$)} &
\colhead{(m~s$^{-1}$)} 
}
\startdata
11409.824 &   -1.54 &  3.35 \\
11705.913 &   25.38 &  3.80 \\
12004.048 &   -8.26 &  3.73 \\
12007.010 &  -10.05 &  4.01 \\
12008.001 &  -11.88 &  3.92 \\
12009.070 &   -2.72 &  3.65 \\
12097.968 &   -8.16 &  3.60 \\
12098.912 &    1.04 &  3.83 \\
12099.844 &    4.90 &  3.63 \\
12100.895 &    4.57 &  3.39 \\
12127.911 &    0.00 &  3.72 \\
12161.808 &    5.87 &  3.89 \\
12189.748 &    2.16 &  4.36 \\
12390.092 &   13.12 &  3.74 \\
12486.759 &   -1.86 &  3.46 \\
12514.756 &   -5.61 &  3.56 \\
12535.722 &   -6.90 &  3.57 \\
12574.690 &    2.29 &  4.02 \\
12712.157 &   -3.63 &  3.71 \\
12777.039 &    7.53 &  3.93 \\
12804.920 &   13.40 &  3.75 \\
12832.961 &   13.60 &  3.91 \\
13179.964 &   -9.51 &  3.82 \\
13195.812 &    0.17 &  3.38 \\
13430.124 &   13.32 &  1.20 \\
13547.904 &   12.84 &  0.99 \\
13842.077 &   -3.95 &  0.97 \\
13934.875 &    9.44 &  0.95 \\
14640.044 &   14.79 &  1.16 \\
14671.928 &   16.81 &  0.72 \\
14673.897 &   25.43 &  0.58 \\
14779.698 &    8.75 &  1.14 \\
14964.087 &    4.24 &  0.96 \\
14985.917 &   -3.91 &  0.96 \\
15014.851 &   -2.75 &  0.98 \\
15015.879 &   -7.25 &  0.97 \\
15019.036 &   -2.05 &  1.51 \\
15041.950 &   -9.03 &  1.05 \\
15042.786 &   -6.49 &  0.97 \\
15043.900 &   -9.16 &  1.00 \\
15048.857 &   -3.67 &  1.28 \\
15075.744 &    4.40 &  0.91 \\
15111.736 &   -2.10 &  1.07
\enddata
\end{deluxetable}

\end{document}